\documentclass[twocolumn,prb,aps,showpacs,superscriptaddress]{revtex4-1}
\usepackage{graphicx}
\usepackage[dvipdfm]{hyperref}
\newcommand{\hm}{h_{mw}}
\newcommand{\vhm}{\vec{h}_{mw}}
\newcommand{\co}{\frac{\gamma}{2}}
\newcommand{\1}{\textcolor{red}}

\usepackage{pst-plot}
\usepackage{epsfig}
\usepackage{pst-grad} 
\usepackage{pst-plot} 
\usepackage{lmodern} 
\newcommand{\mycol}{1}
\newcommand{\beq}{\begin{equation}}
\newcommand{\eeq}{\end{equation}}

\usepackage{amsmath}

\begin{document}
\title{Tunable multi-photon Rabi oscillations in an electronic spin system}

\author{S. Bertaina}
\email{sylvain.bertaina@im2np.fr}
\affiliation{IM2NP-CNRS (UMR 6242) and Universit\'{e} Aix-Marseille, Facult\'{e} des Sciences et Techniques, Avenue Escadrille Normandie Niemen - Case 142, F-13397 Marseille Cedex, France.}
\affiliation{Department of Physics and The National High Magnetic Field Laboratory, Florida State University, Tallahassee, Florida 32310, USA}

\author{N. Groll}
\email{current address: Materials Science Division, Argonne National Laboratory, 9700 S. Cass Avenue, IL 60439, USA.}
\affiliation{Department of Physics and The National High Magnetic Field Laboratory, Florida State University, Tallahassee, Florida 32310, USA}

\author{L. Chen}
\affiliation{Department of Physics and The National High Magnetic Field Laboratory, Florida State University, Tallahassee, Florida 32310, USA}
\affiliation{Department of Chemistry and Chemical Biology, Cornell University, Ithaca, NY 14853-1301, USA}

\author{I. Chiorescu}
\email{ic@magnet.fsu.edu} 
\affiliation{Department of Physics and The National High Magnetic Field Laboratory, Florida State University, Tallahassee, Florida 32310, USA}

\date{submitted}%

\begin{abstract}
We report on multi-photon Rabi oscillations and controlled tuning of a multi-level system at room temperature ($S=5/2$ for Mn$^{2+}$:MgO) in and out of a quasi-harmonic level configuration. The anisotropy is much smaller than the Zeeman splittings, such as the six level scheme shows only a small deviation from an equidistant diagram. This allows us to tune the spin dynamics by either compensating the cubic anisotropy with a precise static field orientation, or by microwave field intensity. Using the rotating frame approximation, the experiments are very well explained by both an analytical model and a generalized numerical model. The calculated multi-photon Rabi frequencies are in excellent agreement with the experimental data.
\end{abstract}

\pacs{03.67.-a 71.70.Ch 75.10.Dg 76.30Da}

\maketitle

\section{introduction}
Harmonic systems are basic manifestation of quantum mechanics and appear in various forms, down to the nanoscale, as electro-magnetic or mechanical oscillators. In the case of a finite number of excited states, multi-level systems are proposed to perform quantum algorithms in size-limited \cite{Leuenberger2001,Leuenberger2002} or scalable \cite{Grace2006} schemes by using microwave (MW) pulses to generate entangled states. The coherent manipulation of spins in a multilevel system is fundamental to the implementation of the Grover algorithm\cite{Grover1997}. At the same time, the quasi-harmonic nature of the system can lead to interesting effects, where the two- or multi-level nature of a system can be interchanged and tuned \cite{Bertaina2009}. 

Spin systems benefit from relatively large coherence and relaxation times, which makes them suitable as quantum bit implementations or as a new type of quantum random access memory \cite{Blencowe2010,Chiorescu2010,Kubo2010,Schuster2010}. In some studies, spin qubits operation is demonstrated at temperatures up to ambient \cite{Akimov2007}. Recent studies demonstrated that in diluted systems, spin-spin dipolar interactions are sufficiently low to allow coherent, quantum manipulations. Such situations are well exemplified by systems like the nitrogen-vacancy color centers in diamonds \cite{Hanson2008,Dutt2007} , N atoms in C$_{60}$\cite{Morley2007}, Er$^{3+}$\cite{Bertaina2007,Bertaina2009a} and Cr$^{5+}$\cite{Nellutla2007} ions and molecular magnets\cite{Bertaina2008,Takahashi2009}.

In a previous work \cite{Bertaina2009}, we demonstrated multi-photon spin coherent manipulation in a multi-level system (Mn$^{2+}$) diluted in MgO, a highly symmetric nonmagnetic matrix. In this work, we report on the possibility of using a multi-level system to tune the nature of quantum Rabi oscillations, using a combination of two parameters: microwave power and/or magnetic field orientation. To this end, one needs a magnetic system with a well defined anisotropy, but sufficiently small in size to be overcome by microwave amplitudes achievable in typical experimental conditions. 

The S=5/2 Mn$^{2+}$ spin is our system of choice. As detailed in the next section, the crystalline anisotropy for this high cubic symmetry is orders of magnitude smaller than the magnetic (Zeeman) energy and therefore the multi-level system is quasi-harmonic. This is essential for successful multi-photon spin manipulation and for state tunability by magnetic field orientation. At the same time, the anisotropy remains smaller than or comparable to the microwave drive, an aspect that is essential for tunability by drive.

In this S=5/2 system, we demonstrate multi-photon Rabi oscillations and controlled tuning of the system in and out of a quasi-harmonic level configuration. Using the rotating frame approximation, the experiments are very well explained by both an analytical model and a generalized numerical model. The calculated multi-photon Rabi frequencies and amplitudes are in excellent agreement with the experimental data.

The article is structured as follows: in Sec.~\ref{sec:spinH} we describe the spin Hamiltonian and its parameters; in Sec.~\ref{sec:expproc} the experimental procedure and setup is detailed; in Sec.~\ref{sec:theo} an analytical and numerical model is given, describing the tunability of the multi-photon dynamics and in Sec.~\ref{sec:expres} we describe the experimental findings.

\section{Quasi-harmonic spin Hamiltonian}\label{sec:spinH}

The spins $S$=5/2 of the Mn$^{2+}$ ions are diluted in a MgO non-magnetic matrix of cubic symmetry $F_{m\bar{3}m}$ (lattice constant 4.216 {\AA}). The Mn$^{2+}$ ions are located in substitutional positions of Mg$^{2+}$ ions. The high degree of symmetry ensures that the spins are seeing an almost isotropic crystalline environment and thus the fourth order magnetic anisotropy can be made much smaller than the Zeeman splittings. Interactions between Mn$^{2+}$ ions and symmetry deformations are neglected. The spin Hamiltonian at resonance is given by \cite{Low1957,Smith1968}:

\begin{eqnarray}\label{eq:1}
    H&=&a/6\left[ {S_x^4+S_y^4+S_z^4-S(S+1)(3S^2-1)/5} \right]\\
&&+\gamma\vec{H}_0\cdot\vec{S}-A\vec{S}\cdot\vec{I}+\gamma\vec{h}_{mw}\cdot\vec{S}
\cos(2\pi ft) \nonumber
\end{eqnarray}
where $\gamma=g\mu_B/h$ is the gyromagnetic ratio ($g=2.0014$ the $g$-factor, $\mu_B-$ Bohr's magneton and $h-$ Planck's constant), $S_{x,y,z}$ are the spin projection operators, $\vec{S}$ is the total spin, $a =$ 55.7~MHz is the anisotropy constant, $A = 244$~MHz is the hyperfine constant of $^{55}$Mn ($I = 5/2$), $h_{mw}$ and $f$ represent the MW amplitude and frequency respectively, and $\vec{H}_0$ is the static field ($\vec{H}_0\perp \vec{h}_{mw}$). In our experiments, the static field ensures a Zeeman splitting of $\gamma H_0\approx f\sim$9~Ghz, much stronger than all other interactions of Eq.(\ref{eq:1}). This implies that (i) $\vec{H}_0$'s direction can be approximated as the quantization axis and (ii) coherent MW driving is confined between levels of same nuclear spin projection $m_I$ (see also refs. \onlinecite{Bertaina2009,Hicke2007,Schweiger2001}).

\begin{figure}
\psset{unit=0.07\columnwidth}
\includegraphics[bb=142 452 392 686, clip, width=\columnwidth]{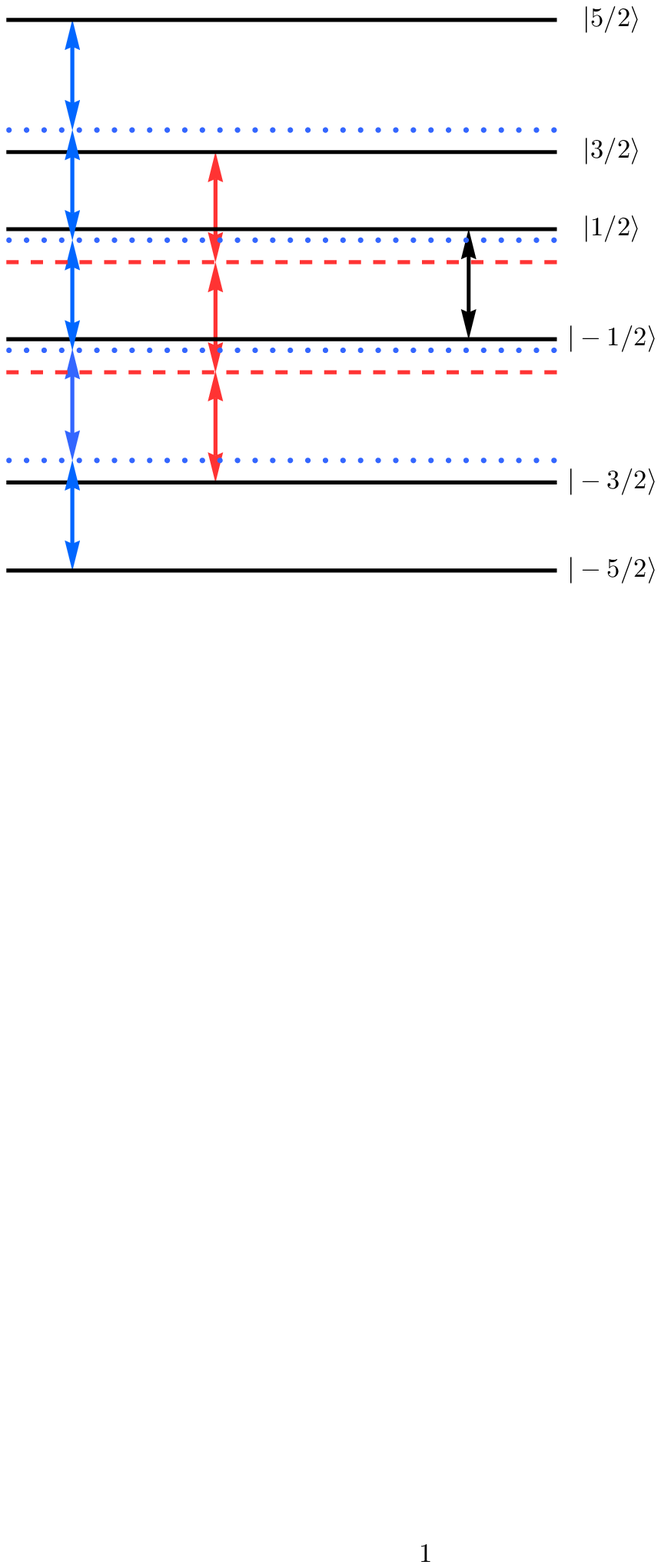}
\caption{Level diagram of Mn$^{2+}$ spin in a cubic crystal field for a Zeeman splitting $\gamma H_0=E_{1/2}-E_{-1/2}$. Arrows indicate one photon (in black), three (in red) and five (in blue) photon monochromatic transitions, between spin projections $m$ as shown on the left side. The dashed and dotted lines show equidistant virtual levels enhancing the multi-photon transitions. }
\label{fig:level}
\end{figure}

The six-level system is consequently quasi-harmonic, as shown by the level diagram in Fig.~\ref{fig:level}. The dashed and dotted lines show virtual levels of equidistant separation $g\mu_BH_0=hf$, which strongly enhance the multi-photon transition probabilities. Since $a\ll hf$, the distance between the real and virtual levels in Fig.~\ref{fig:level} is exaggerated for clarity. The number of arrows for each monochromatic transition suggest the type of multi-photon coherent excitation (one, three or five-photon). A coherent Rabi superposition of the $m=-5/2\dots 5/2$ states (as counted on the left side of the figure) can thus be obtained, strongly dependent on the microwave drive power and nutation time.

\section{Experimental procedure}\label{sec:expproc}

\begin{figure}[h]
\includegraphics[bb=143 583 481 744,clip, width=\columnwidth]{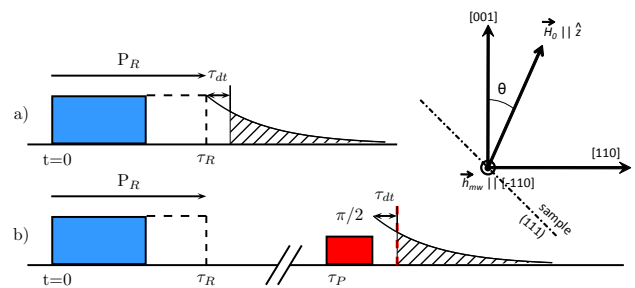}
\caption{(Color online) Pulse sequence used in the multi-photon Rabi oscillation measurements, which starts with a strong excitation pulse P$_R$ inducing the multi-photon nutation. a) Right after the pulse P$_R$, the FID signal gives the transverse magnetization state between -1/2 and 1/2. b) After a time $\tau_P\gg T_2$, a $\pi/2$ pulse rotates the longitudinal magnetization so it can be probed and so is the density of states in levels $|-1/2\rangle$ and $|1/2\rangle$. (see text for more details). The insert shows the static and electromagnetic fields orientations by respect to the crystalline axes.}
\label{fig:seq1}   
\end{figure}

Rabi oscillations measurements have been performed in a Bruker Elexsys 680 pulse EPR spectrometer working at about $f=9.6$~GHz ($X$-band). The sample is a 3x3x1~mm${^3}$ single crystal of MgO doped by a small amount of Mn$^{2+}$ (about few ppm). The orientation between the sample and the static field is controlled by a goniometer (precision 1$^\circ$) with the rotation axis parallel to the microwave magnetic field direction $\vhm$. In the measurements presented here, $\vhm||[-110]$ (see Fig.~\ref{fig:seq1} insert). All measurements have been made at room temperature. Calibration of the microwave field $\hm$ have been made using the BDPA standard: a small amount of BDPA containing isotropic S=1/2 spins give a Rabi frequency of exactly $F_R=\mu_B \hm/h$.

By design, the detection is sensitive to frequency $f$ and therefore it can probe only transitions between consecutive spin projections, $m\leftrightarrow m+1$. As the transition $|-1/2\rangle\leftrightarrow|1/2\rangle$ is of highest probability, as ensured by Fermi's golden rule, and is not sensitive to crystal strain effects we choose to use the levels $m=\pm 1/2$ as a probe of the six-level dynamics.

By definition, the magnetization is given by $\langle S_i \rangle\equiv \mathrm{Tr}(\rho S_i)$ where $\rho$ is the  density matrix and the $S_i$ is the S=5/2 spin operator $i=x,y,z$. Since we probe only the transition $|-1/2\rangle\leftrightarrow|1/2\rangle$ we define the magnetization of the subset  $\langle S_i \rangle_{1/2}\equiv \mathrm{Tr}(\rho_{1/2} S_i)$ where $\rho_{1/2}$ is the central block matrix of $\rho$ corresponding to levels $\pm 1/2$.

The drive and detection of the multi-photon Rabi oscillations are implemented with the pulse sequence presented in Fig.~\ref{fig:seq1}. A drive pulse P$_R$ of duration $\tau_R$ and resonant to the $|-1/2\rangle \leftrightarrow |+1/2\rangle$ transition, is applied at $t=0$. At the end of this pulse, the density of states has been coherently changed.

Right after the $P_R$ pulse, spin dephasing and line inhomogeneity induce a free induction decay (FID) signal\cite{Bloch1946}. The first point of this FID (at t=$\tau_R$) gives the transverse magnetization traced on the $|\pm 1/2\rangle$ subset: $\langle S_x\rangle_{1/2}=\rho_{-1/2,1/2}+\rho_{1/2,-1/2}$. Unfortunately, the dead time $\tau_{dt}$ of the spectrometer ($\sim 80$~ns) prevents the measurement of this first point. If the entire inhomogenous line of width $\Delta H_0$ is excited by a sufficiently short P$_R$ pulse (non selective pulse, here $\tau_R^{-1}>\gamma \Delta H_0/2\pi=0.27$~MHz), the FID is simply the Fourier transform of the absorption line and is an exponential decay in MgO:Mn$^{2+}$. Therefore, we integrate the detected FID and the result is proportional to its first point. Note that for a broad line (not our case), as often seen in solid state paramagnetic systems, the FID oscillates and the first point cannot be found so easily\cite{Schenzle1980,Kunitomo1982}.
 
The pulse sequence in Fig.~\ref{fig:seq1}(a) detects the coherent evolution of the transverse magnetization $\langle S_x\rangle_{1/2}$, as a function of pulse length $\tau_R$. To probe the coherent evolution of the longitudinal magnetization $\langle S_z\rangle_{1/2}=\rho_{1/2,1/2} -\rho_{-1/2,-1/2}$ (also traced on the $|\pm 1/2\rangle$ subset), we use the sequence shown in Fig.~\ref{fig:seq1}(b). After the Rabi pulse P$_R$, one waits a time $\tau_P-\tau_R$, smaller than system's relaxation time $T_1$ but larger than the decoherence time $T_2$. Thus, at $t=\tau_P$, only the longitudinal magnetization is non-zero \cite{Stoll1998}.

A $\pi/2$ pulse follows, rotating the $|\pm 1/2\rangle$ population in a coherent mixture, which is located in the $xy$ plane (or the ``equatorial plane''). The detection observes the free induction decay (FID) towards zero of this mixture, giving essential information such as: the initial total magnetization in the $xy$ plane, its FID decoherence time ($T_2^*$) and potential shift away from the resonance static field.

As indicated with arrows in Fig.~\ref{fig:level}, in a cubic symmetry the three and five photon transitions do use the $|\pm 1/2\rangle$ states as intermediate ones. Therefore, the $\pi/2$ pulse can distinguish between the various multi-photon Rabi oscillations, due to their different frequencies as detailed in the theoretical section below.

\section{Rabi rotations in a tunable multi-level system. Theoretical treatment.}\label{sec:theo}

The aim of this section is to present a model which describes the coherent multiphoton Rabi oscillation that occur in a quasi-harmonic large spin system under  monochromatic radiation. To compute the multiphoton Rabi frequencies, we used the density matrix theory\cite{Brewer1975} applied to the Mn$^{2+}$ ion ($S$=5/2, $I$=5/2) in the rotating frame approximation.

\subsection{Analytical calculation}\label{sec:theo:anali}
Let us consider a quantum system with six states $| S_z\rangle$, $S_z$=$\left\{-5/2, -3/2, -1/2, 1/2, 3/2, 5/2 \right\}$, irradiated by an electromagnetic field in resonance with the $-1/2$ and 1/2 levels. We assume initially $I=0$, but we will describe the effect of the nuclear spin at the end of this section. The Hamiltonian of the system is :
\begin{equation}\label{eq:2}
    \mathcal{H}=\hat{E}+\hat{V}(t)=\sum_{S_z=-5/2}^{5/2} E_{S_z}|S_z\rangle\langle S_z|+\hat{V}(t),
\end{equation}
with $E_{Sz}$ the static energy levels, $\hat{V}(t)=\frac{\gamma}{2}h_{mw}(\hat{S}_++\hat{S}_-)\cos\left(2\pi ft\right)$, $S_+/S_-$ the raising/lowering operators and $f=E_{1/2}-E_{-1/2}$.

\begin{figure}
\includegraphics[bb=0 0 187 132,width=\mycol\columnwidth]{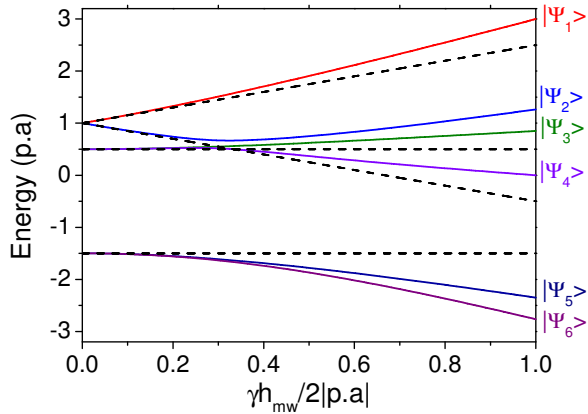}
\caption{(color online) Eigenvalues of $\mathcal{H}_{RWA}$ as a function of $\tilde h$. The dashed lines are the limit $\hm\ll p.a$.} \label{fig:dressed}
\end{figure}

In a cubic symmetry and in first order in perturbation ($a \ll H_0$), the static energy levels are given by:\\
\begin{eqnarray}
   \nonumber 
	E_{\pm 5/2}&=&\pm 5/2 \gamma H_0+1/2 p a+\mathcal{O}(a^2)\\ 
	E_{\pm 3/2}&=&\pm 3/2 \gamma H_0-3/2 p a+\mathcal{O}(a^2)\\
	E_{\pm 1/2}&=&\pm 1/2 \gamma H_0+ p a+\mathcal{O}(a^2) \nonumber
\end{eqnarray}
where\cite{Low1957} $p=1-5(A^2B^2+B^2C^2+C^2A^2)$, with $A,B$ and $C$ representing the cosine directors of $\vec{H}_0$ with the crystalline axes. For our geometry, $p$ is given by:

\begin{equation}\label{eq:p}
    p=1-5\sin^2\theta+\frac{15}{4}\sin^4\theta,
\end{equation}
with $\theta$ the angle between $\vec{H}_0$ and the $c$ axis $[001]$. Since $H_0\gg \hm$, the negative frequency part of $V(t)$ responsible of the Bloch-Siegert shift\cite{Bloch1940} can be neglected and we can use the rotating wave approximation (RWA) to make Eq.~(\ref{eq:2}) to be time independent. Since in a conventional pulsed spectrometer we have access only to the first harmonic detection ($hf$), we define the unitary transformation $U(t)=\exp(-i2\pi f \hat{S}_z t)$ and apply it to Eq. (\ref{eq:2}).
In the rotating frame, the Hamiltonian (\ref{eq:2}) becomes \cite{Hicke2007, Leuenberger2003} :
\begin{equation}
\label{eq:H_rwa}
\mathcal{H}_{RWA}=U\mathcal{H} U^\dag+i\hbar\frac{\partial U}{\partial t}U^\dag= 
\end{equation}
\begin{equation}
=\left(\begin{matrix}
\frac{1}{2}pa  & \frac{\sqrt{5}}{2} V & 0 & 0 & 0 & 0\\
\frac{\sqrt{5}}{2} V&-\frac{3}{2} p a & \sqrt{2} V  & 0 & 0 & 0\\
0 &\sqrt{2} V & pa & \frac{3}{2}V & 0 & 0\\
0 &  0  & \frac{3}{2}V  & pa  & \sqrt{2} V  & 0\\
0 &  0  & 0 &  \sqrt{2} V  & -\frac{3}{2} p a  & \frac{\sqrt{5}}{2} V\\
0 &0 & 0 & 0 &  \frac{\sqrt{5}}{2} V  & \frac{1}{2}pa \\
\end{matrix}\right)
\label{eq:H_rwa_noI}
\end{equation}
where $V=\gamma\hm/2$. By diagonalization, the eigenenergies $E_n/pa$ of the dressed states $|\Psi_n\rangle$ are calculated as a function of $\tilde{h}\equiv V/pa$, as illustrated in Fig.~\ref{fig:dressed} (for $p>0$).

For high anharmonicity or low microwave power $\tilde{h}\rightarrow 0$, in which case $|\Psi_1\rangle=(|1/2\rangle+|-1/2\rangle)/\sqrt{2}$ and $|\Psi_2\rangle=(|1/2\rangle-|-1/2\rangle)/\sqrt{2}$ are the coherent superposition of states $|\pm 1/2\rangle$. The other wavefunctions remain unchanged  $|\Psi_3\rangle=|5/2\rangle$, $|\Psi_4\rangle=|-5/2\rangle$, $|\Psi_5\rangle=|3/2\rangle$ and $|\Psi_6\rangle=|-3/2\rangle$. 

The splitting between $|\Psi_1\rangle$ and $|\Psi_2\rangle$ is detected as an oscillation of frequency $F_R=E_2-E_1=3\co \hm$, in agreement with the general formula \cite{Schweiger2001}:

\begin{equation}\label{eq:Rabi1Photon}
    F_R^{(1)}=\co\hm\sqrt{S(S+1)-S_z(S_z+1)}
\end{equation}
for one photon Rabi oscillations between consecutive states $S_z$ and $S_z+1$.

When the microwave power $\hm$ increases, the degeneracy of $|\Psi_n\rangle$ ($n=$3,4,5,6) is lifted, allowing coherent three and five photon transitions ($|-3/2\rangle\leftrightarrow|3/2\rangle$ and $|-5/2\rangle\leftrightarrow|5/2\rangle$ respectively). Note that two and four photon processes are out of resonance and therefore not discussed here. In first order perturbation in  $\tilde{h}$, only the one photon process exists between states $|\Psi_1\rangle$ and $|\Psi_2\rangle$. In third order perturbation, a three photon process mixes the states $|\pm 3/2\rangle$, leading to a pure 3-photon Rabi oscillation with frequency $F_R^{(3)}= \frac{24}{25} \tilde{h}^3 pa$, as detailed in the Appendix. This three photon process perturbs the  aforementioned one photon process which now has the Rabi frequency $F_R^{(1')}=F_R^{(1)}-\frac{24}{25}\tilde{h}^3 pa$. 

In fifth order perturbation, the  five photon process mixes the states $|\pm 5/2\rangle$, leading to a Rabi frequency $F_R^{(5)}=\frac{15}{2} \tilde{h}^5 pa$ (pure 5-photon process). Similarly, the five photon process will change the dynamics of the three and one photon Rabi oscillations (see Appendix~\ref{appendix}, Eqs.~\ref{eqn:F1}-\ref{eqn:F5}).  

For $\tilde{h}\rightarrow \infty$, the microwave field excites all transitions and the six levels are coherently driven by a 5-photon process. The dynamics of such a system becomes that of an isotropic spin S=1/2 . The diagonal of the Hamiltonian (\ref{eq:H_rwa_noI}) becomes negligible and the diagonalization simply rotates the Hamiltonian along the $h_{mw}$ field.\\

When the nuclear spin is included, the model becomes more complicated but analytical solutions can still be found. The full spin Hamiltonian of $^{55}$Mn$^{2+}$ ($S=5/2$, $I=5/2$) is a 36 x 36 matrix. If we assume no forbidden nuclear transitions ($\Delta m_I=0$, with $m_I$ quantifying the $I$ projections), the Hamiltonian (\ref{eq:1}) can be separated into six 6x6 matrices of form Eq. (\ref{eq:2}).  

The static energies $E_{S_z}$  now depend also on $m_I$ and $A$ and have been computed in first order perturbation in $A/H_0$  by Low~\cite{Low1957}. Having the new $E_{S_z}$ renormalized by the hyperfine field, we can use the same procedure as above to compute the eigenvalues of the dressed states and the Rabi splittings.    

Finally, note that all other transitions can be computed by the method presented here, by changing the value of $f$ in Eq.~(\ref{eq:2}). For instance, for $f=(E_{-3/2}-E_{1/2})/2$ one can study the 2-photon Rabi oscillations in the transition $|-3/2\rangle\leftrightarrow|1/2\rangle$).

\subsection{Numerical calculation}\label{sec:theo:numerical}

The analytical model presented above describes the multiphoton frequencies of the Rabi oscillations very well but fails to describe their intensities (the time evolution of spin populations). In order to describe the experimental results we developed a model using Hamiltonian~(\ref{eq:1}), which shows the actual crystal field parameters and includes the hyperfine coupling between the $S=5/2$ electronic spin and the $I=5/2$ nuclear spin of $^{55}$Mn$^{2+}$. 

The anisotropy $a$ and hyperfine $A$ constants are two orders of magnitude smaller than the electron Zeeman interaction. Therefore, it is appropriate to assume that the orientation of the static field imposes the quantization axis $\tilde{z}$. In this base, Hamiltonian (\ref{eq:1}) rewrites as in (\ref{eq:2}), where $E_{S_z}$ are obtained by exact diagonalization and reflect the hyperfine coupling.

\begin{figure}
\includegraphics[bb=30 7 796 544,width=\mycol\columnwidth]{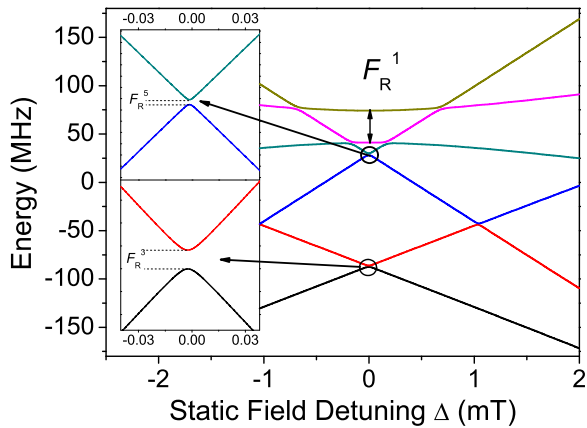}
\caption{(color online) Dressed states energies for $H_0||[1 0 0]$, $h_{mw}=0.8$~mT and $m_I=1/2$. The large anticrossing $F_R^1$ is the 1-photon Rabi splitting. The insets are close-ups of the 3-photon and 5-photon Rabi splittings.} \label{fig:RWA}
\end{figure}

Once we get the form ($\ref{eq:2}$) for $\mathcal{H}$, the transformation in the rotating frame become obvious. Experimentally, the spectrometer signal represents the transient magnetic resonant signal in the rotating frame, obtained by mixing the signal reflected by the cavity ($S(t)\cos(2\pi ft)$) and the reference arm ($\cos(2\pi ft)$). This leads to a frequency-independent signal $S(t)$ and a double frequency signal (which is filtered out). The unitary transformation is fixed by the mw-frequency $f$, with $U(t)\exp(-i2\pi\hat{S}_zft)$ and using Eq.~(\ref{eq:H_rwa}) one gets:
\begin{equation}\label{eq:5}
    \mathcal{H}_{RWA}=\co\hm \hat{S}_x+\hat{E}-f\hat{S}_z
\end{equation}
For a fixed $m_I$, the static field $H_0$ is chosen to satisfy the condition $E_{1/2}-E_{-1/2}+\Delta=f$ where $\Delta$ represents the detuning of the static field, away from resonance. As an example, the eigenvalues of Hamiltonian (\ref{eq:5}), for $H_0||[1 0 0]$ and nuclear projection $m_I=1/2$, are shown in Fig.~\ref{fig:RWA}, as a function of field detuning $\Delta$. In these simulations, the $h_{mw}=0.8$~mT, sufficiently strong to induce Rabi splitting with three and five photons, as shown in insets.             
  
The dynamics of Hamiltonian~(\ref{eq:5}) is described by the time evolution of the 6x6 density matrix $\rho(t)$:
\begin{equation}\label{eq:6}
    i\frac{d\rho}{dt}=[\mathcal{H}_{RWA}, \rho].
\end{equation}
In the rotating frame (Dirac or interaction picture), the solution is $\rho(t)=U_p(t)\rho_0 U^\dag_p(t)$ with $\rho_0$ the matrix density at thermal equilibrium :
\begin{equation}\label{eq:7}
    \rho_0=\frac{\exp(-\hat{E}/kT)}{Tr (\exp(-\hat{E}/kT))}
\end{equation}  
and $U_p(t)$ is the propagator operator: $U_p(t)=\exp(-i2\pi \mathcal{H}_{RWA} t )$.

When the Rabi pulse $P_R$ is applied, the spin population is coherently manipulated. At the end of the sequence a weak $\pi/2$ pulse will selectively probe the difference of population between states -1/2 and +1/2. The signal out of the spectrometer is $S(t)=\sigma_{-1/2}(t)-\sigma_{1/2}(t)$. Note that for a spin $S=1/2$, $S(t)=\langle S_z\rangle(t)$

\section{Experimental results}\label{sec:expres}

A typical MW absorbtion spectrum of MgO:Mn$^{2+}$ as a function of $H_0$, shows 6 sets (one for each $m_I$ projection) of 5 lines ($S_z\rightarrow S_z+1, S_z=-5/2\dots 3/2$). In the experiments described below, the magnetic field is tuned to study resonances in the set corresponding to $m_I=1/2$. All experimental and theoretical results are valid to any other nuclear spin projection.

\subsection{Tunable multiphoton Rabi oscillations by changing the orientation of the static field.}

\begin{figure}
\includegraphics[bb=10 1 192 133,width=\mycol\columnwidth]{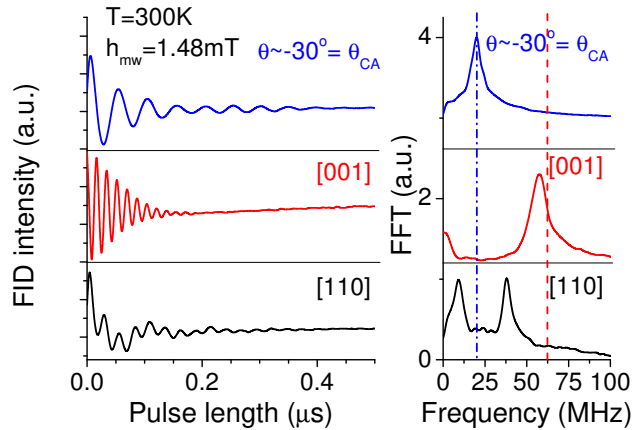}
\caption{(color online) Rabi oscillations and its Fourier transforms for three $H_0$ orientations and for a microwave field $h_{mw}$=1.48~mT. For $H_0$ along the compensation angle (blue curves), the dynamics are the same as that of a $S=1/2$ system. For H$_0||[001]$ (red) the microwave field is smaller than the effective anisotropy and only the 1-photon process is induced. For H$_0||[110]$ (black) the microwave field and the effective anisotropy are comparable and the 3-photon process becomes visible.   } \label{fig:3Rabi}
\end{figure}

The static field orientation is described by the $p$ parameter (see Eq.~\ref{eq:p}) and therefore can tune the reduced microwave field $\tilde{h}=V/(pa)$. We measured Rabi oscillations and corresponding Fourier transforms at constant microwave field ($h_{mw}$=1.48~mT), for three different orientations (and $p$ values) of $H_0$, as shown in Fig.~\ref{fig:3Rabi}. One can see a highly anisotropic behavior of the coherent dynamics depending on the relative orientation between $H_0$ and crystal axes.

For $H_0 || [001]$ ($p=1$, maximum value), only one frequency is observed but its value (57~MHz) is smaller than the one expected for the 1-photon Rabi frequency in the transition $-1/2\leftrightarrow 1/2$ in S=5/2 (62.5 MHz, see Eq.~(\ref{eq:Rabi1Photon}) and red dashed line in Fig. \ref{fig:3Rabi}). Although the multi-photon Rabi oscillation is not clearly resolved, it indirectly slowing down the expected 1-photon dynamics, as described in Sec.~\ref{sec:theo:anali} by the difference $F_R^{(1)}-F_R^{(1')}=F_R^{(3)}$.

For $H_0||[110]$ ($p=-1/4$), in addition to the high frequency 1-photon process, a low frequency 3-photon Rabi oscillation becomes observable, due to a lower $p$ value and thus a larger $\tilde{h}$. Finally, for $\theta=\theta_{CA}$ (the compensation angle of the cubic anisotropy: $p(\theta_{CA})=0$) and hence $\tilde{h}\rightarrow \infty$, only one Rabi frequency is observed (Fig.~\ref{fig:3Rabi} top). Its value is the same as the one expected for an isotropic spin 1/2 (dashed-dot line in fig \ref{fig:3Rabi}). As explained in Sec.~\ref{sec:theo:anali} and Appendix, a coherently driven 5-photon process (and only this process) is induced, leading to a S=1/2 spin dynamics.

\begin{figure}
\includegraphics[bb=0 0 182 133,width=\mycol\columnwidth]{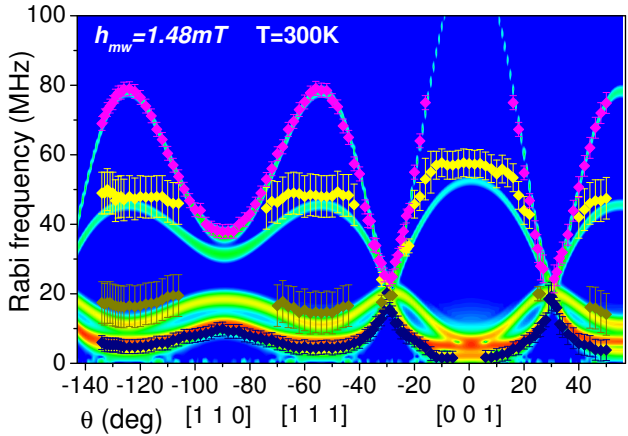}
\caption{(color online) Rabi frequency distribution of MgO:Mn$^{2+}$ while the static field H$_0$ rotates in the plan (-110) and for a microwave field $h_{mw}$=1.48mT. H$_0$=348.5~mT and $f$=9.676GHz are set to probe the transition $|m_S=-1/2,m_I=1/2> \leftrightarrow |m_S=1/2,m_I=1/2>$.  The points come from the Fourier transform of experimental data. The error bars are the width of the Fourier transform (consequence of the damping). The contour plot is calculated with the numerical model given in Sec.~\ref{sec:theo:numerical}.} \label{fig:rabiangle2d}
\end{figure}

A detailed $\theta$ dependence of detected Rabi frequencies, found by fast Fourier transformation (FFT) of coherent oscillations, is given in Fig.~\ref{fig:rabiangle2d}, for the same value of $\hm$. The error bars are FFT linewidths and are due to Rabi decay processes. The contourplot is computed using the numerical model described in Sec.\ref{sec:theo:numerical}, with no adjustable parameters: the crystal field and hyperfine constant are from Ref.~\onlinecite{Low1957} and the MW field has been independently calibrated using a standard. The agreement between our model and the experimental data is quite good. However there are some discrepancies when $|pa|$ is small (but not zero) compared to $\gamma\hm/2$. In this case, the 4 EPR satellite lines are close to the central line and can overlap due to small inhomogeneity caused by crystal field strain. Since the multi-photon Rabi frequency is highly dependent on $|pa|$ (specially when $|pa|$ is small), the distribution in $pa$ induces a high distribution of multi-photon Rabi frequencies, which in turn overdamps the Rabi oscillation signal.

\subsection{Tunable multiphoton Rabi oscillations by changing the MW field.}

\begin{figure}
\includegraphics[bb=14 1 174 124,width=\mycol\columnwidth]{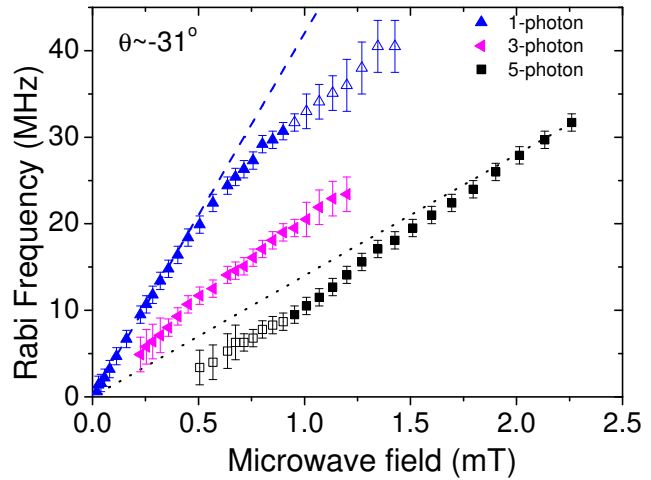}
\caption{(color online) Rabi frequencies of Mn$^{2+}$ spin, as a function of drive field, obtained by FFT of detected coherent oscillations. The error bars are the FFT linewidths. The dashed line indicates the expected 1-photon dependence given by Eq.~\ref{eq:Rabi1Photon}, whereas the dotted line models the isotropic $S=\frac{5}{2}$ case (see Eq.~\ref{eq:A7}).  } \label{fig:power}
\end{figure}
Aside the tuning method based on static field orientation, described in the previous section, an alternative method is based on MW drive intensity. To demonstrate the method, we apply a static field not far from the compensation angle $\theta_{CA}$ described above, such that the spin system is slightly anisotropic. We chose an angle $\theta=31\pm 1^{\circ}$ and record coherent Rabi oscillations as a function of MW field $h_{mw}$. The FFT peaks (Rabi frequencies) are shown in Fig.~\ref{fig:power} by points and the error bars represent the FFT linewidths. The dashed blue line is the MW field dependence of Rabi frequency, for a pure 1-photon transition between levels $|-1/2\rangle \leftrightarrow |1/2\rangle$ of the $S=5/2$ system ($F_R=3\co \hm$, see eq. \ref{eq:Rabi1Photon}). The dotted black line is the MW field dependence of Rabi frequency in the case of a pure isotropic spin ($F_R=\co \hm$, see Eq.~\ref{eq:A7}).    

Indeed, at low MW power, the system has the dynamics of an anisotropic S=5/2 system, as shown by the very good agreement between experimental points (up triangles) and the expected dependence for a 1-photon process (dashed line). At intermediate powers, the FFT clearly shows the existence of a three photon process (left triangles), as well as of a 5-photon process. The later is indicated by empty squares for shallow peaks in the FFT and by full squares, corresponding to well defined FFT peaks.


A quantitative numerical description is quite difficult because of the extreme non linearity of $\tilde{h}$ near $\theta_{CA}$. At the compensation angle ($\theta_{CA}=29.67^\circ$) $\tilde{h}\rightarrow\infty$, but a small change in $\theta$ induces significant changes in $\tilde{h}$ (e.g. at $\theta=30^{\circ}$, $\tilde{h}=16.3 h_{mw}$, at $\theta=31^{\circ}$, $\tilde{h}=4.1 h_{mw}$ and at $\theta=32^{\circ}$, $\tilde{h}=2.4 h_{mw}$). As the accuracy of our goniometer is $1^\circ$, a numerical quantitative description around $\theta_{CA}$ is not relevant. Qualitatively, one notes that for $h_{mw}>0.5$~mT, the one-photon Rabi frequency deviates from $F_R^{(1)}$ of Eq.~\ref{eq:Rabi1Photon}, due to the formation of multi-photon processes. For high microwave powers ($h_{mw}> 1$~mT), the one-photon FFT peak is barely visible (shown by empty up triangles) while the five photon process dominates entirely the dynamics. This is indicated by the good agreement between the 5-photon FFT peaks (filled squares) and the dotted line resulting from an isotropic $S=5/2$ model (or a spin 1/2), as explained by Eq.~\ref{eq:A7}. 

Our experiment demonstrates that a sufficiently intense microwave field can tune the spin dynamics from that of an anisotropic (unequidistant) system, to that of an equidistant, quasi-harmonic 5-level system (equivalent to a two-level system).

\subsection{Five-photon Rabi oscillations at the compensation angle.}

\begin{figure}
\includegraphics[bb=0 0 181 126,width=\mycol\columnwidth]{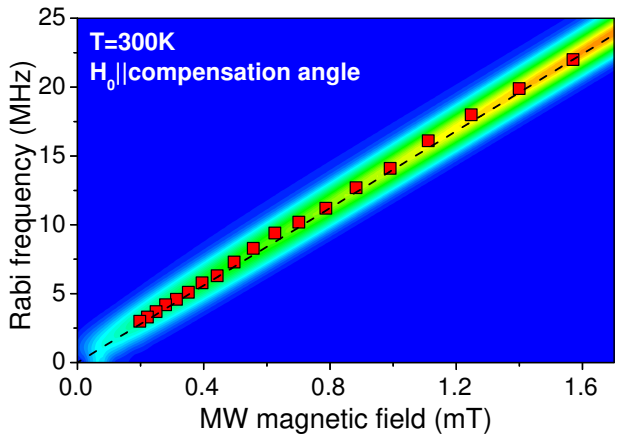}
\caption{(color online) Rabi frequencies as FFT peaks (filled squares) of coherent oscillations for $H_0$ along the compensation direction of the cubic anisotropy, $\theta=\theta_{CA}$. The contour plot is numerically computed, as described in Sec.~\ref{sec:theo:numerical}, while the dashed line shows the expected behavior for an isotropic $S=5/2$ system (see Eq.~\ref{eq:A7}).  }
\label{fig:compens}
\end{figure}

When the static field is applied along a direction for which $\theta=\theta_{CA}$ (that is, $\theta=29.67^{\circ}$), the cubic anisotropy is geometrically compensated and therefore $p=0$ and $\tilde{h}\rightarrow\infty$. As explained in Sec.~\ref{sec:theo:anali} and Appendix, one expects a perfectly isotropic $S=5/2$ spin system.

We performed room temperature Rabi oscillations for this field orientation. The FFT of the coherent oscillations are single-peaked, as shown in Fig.~\ref{fig:compens} with filled squares (the FFT width is smaller than symbol's size). The contour plot is calculated using the numerical procedure described in Sec.~\ref{sec:theo:numerical}, whereas the dashed line indicates the analytical result of Eq.~\ref{eq:A7}. One observes an excellent agreement between the analytical and numerical models and the experimental data points (no fit parameter has been used).

During a Rabi rotation for this isotropic case, the spin density gradually moves upward on the equidistant 6-level ladder and, when one reaches the highest level ($S_z=5/2$), the dynamics is reversed. In contrast to a one-photon dynamics, here we have a collective five photon process, ensuring a continuous nutation within the $S_z=-5/2\ldots 5/2$ space.

\section{Discussion and conclusion.}
The exerimental data and the model presented here are using a monochromatic source to both induce and probe the multi-photon Rabi oscillations. Therefore, we restrict the types of multi-photon transitions that are measurable, to those which exist at the same resonance static field as the one photon transition (used to induce and probe). In order to probe directly all types of multi-quanta transitions, two tunable frequencies will be needed: one to induce the multi-photon oscillations (i.e. to induce a 2-photon Rabi oscillation:  $f_1=(E_{1/2}-E_{-3/2})/2$ and one to probe it $f_2=E_{1/2}-E_{-1/2}$). Note that the experiments described here, with a single or two tunable frequencies, have the appeal of being applicable to on-chip studies, using magnetic detectors placed in strong magnetic fields\cite{Chen2010,Groll2010,Groll2009}. 

In conclusion, we report on single and multi-photon coherent rotations in a $S=5/2$ spin system featuring a cubic anisotropy. The anisotropy is much smaller than the Zeeman splittings, such as the six level scheme shows only a small deviation from an equidistant diagram. This allows us to tune the spin dynamics by either compensating the cubic anisotropy with a precise static field orientation, or by microwave field intensity. In both cases, we can transit the system between an anisotropic to an isotropic situation, the later case showing a spin dynamics corresponding to a two-level system. The experimental data is well explained by both a theoretical and numerical model.

\acknowledgments
This work was supported by City of Marseille, Aix-Marseille University (BQR grant), NSF Cooperative Agreement Grant No. DMR-0654118, NSF grants No. DMR-0645408, and the Sloan Foundation. We thanks A. Verga for fruitful discussion, G. Gerbaud (BIP-UPR9036) and the pluridisciplinary EPR pole of Marseille for technical support.

\appendix
\section{Eigenvalues of dressed states}\label{appendix}
Diagonalization of Hamiltonian (\ref{eq:H_rwa_noI}) can be achieved analytically. Since the most relevant parameter is the value of the microwave field compared to the anharmonicity parameter $pa$, we define the reduced field $\tilde{h}=\gamma\hm/(2pa)$. The eigenvalues of (\ref{eq:H_rwa_noI}) are the solutions of the polynomial equations :
\begin{eqnarray}\label{eq:A1}
    \nonumber \mathcal{P}_\pm&=&6\pm 9\tilde{h}+18\tilde{h}^2 \pm 15\tilde{h}^3 \dots\\&& +(-14\mp 12\tilde{h}-26 \tilde{h}^2)\mathcal{E}
 \mp 12\tilde{h}\mathcal{E}^2+8\mathcal{E}^3=0
\end{eqnarray} 
\\
(i) For $\tilde{h}\rightarrow 0$, in first-order perturbation, equation (\ref{eq:A1}) becomes :
\begin{equation}\label{eq:A2}
    \mathcal{P}_\pm\approx 6\pm 9\tilde{h}+(-14\mp 12\tilde{h})\mathcal{E}
 \mp 12\tilde{h}\mathcal{E}^2+8\mathcal{E}^3=0
\end{equation} 
with solutions $\mathcal{E}_{1,2}=1\pm 1.5\tilde{h}$, $\mathcal{E}_{3,4}=0.5$ and $\mathcal{E}_{5,6}=-1.5$. The RWA wavefunctions are now : $|\Psi_1\rangle=(|1/2\rangle+|-1/2\rangle)/\sqrt{2}$, $|\Psi_2\rangle=(|1/2\rangle-|-1/2\rangle)/\sqrt{2}$, $|\Psi_3\rangle=|5/2\rangle$, $|\Psi_4\rangle=|-5/2\rangle$, $|\Psi_5\rangle=|3/2\rangle$ and $|\Psi_6\rangle=|-3/2\rangle$. Rabi oscillations occur only between states $|\Psi_1\rangle$ and $|\Psi_2\rangle$ with frequency: $F^1_R=(\mathcal{E}_1-\mathcal{E}_2)pa$ (see Eq.~\ref{eq:Rabi1Photon}). We found the classical result that during the resonance, $|1/2\rangle$ and $|-1/2\rangle$ are mixed by the EM-wave and the other states stay unperturbed.

(ii) When $\tilde{h}$ is increased, we computed the solutions of (\ref{eq:A1}) up to the fifth-order perturbation in $\tilde{h}$. The Rabi frequencies $F_R=(\mathcal{E}_+-\mathcal{E}_-)|pa|$ are:
\begin{eqnarray}
    F_R^1&=& pa \left(3\tilde{h}-\frac{24\tilde{h}^3}{25}-\frac{768\tilde{h}^5}{125}\right)\label{eqn:F1},\\
	 F_R^3&=& pa \left(\frac{24\tilde{h}^3}{25}-\frac{339\tilde{h}^5}{250}\right)\label{eqn:F3},\\
	 F_R^5&=& pa \left(\frac{15\tilde{h}^5}{2}\label{eqn:F5}\right).
\end{eqnarray}

\begin{figure}
\includegraphics[bb=0 0 173 127,width=\mycol\columnwidth]{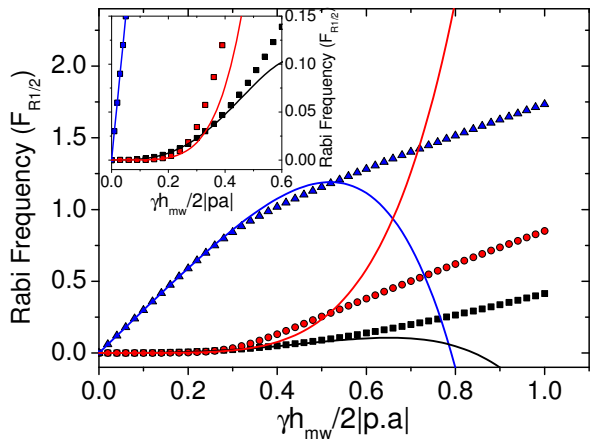}
\caption{(color online) Rabi frequency dependence as a function of the reduced MW-field $\tilde{h}$. Data points are numerically computed by solving Eq.~\ref{eq:A1}, while lines are the analytical perturbation developments (\ref{eqn:F1})-(\ref{eqn:F5}). Inset shows a close-up at low fields/low Rabi frequencies.} \label{fig:A1}
\end{figure}

These equations are used to simulate the continuous lines in Fig.~\ref{fig:A1}, which are compared to the numerical solutions of Eq.~\ref{eq:A1} (discrete data points). The inset shows a close-up view at low microwave powers and Rabi frequencies. One can conclude that for powers up to $\tilde{h}\sim 1/2$, the perturbative analysis works very well.

(iii) When $\tilde{h}\rightarrow \infty$, Eq.~(\ref{eq:A1}) becomes:
\begin{equation}\label{eq:A3}
    \mathcal{P}_\pm\approx \pm 15\tilde{h}^3-26\tilde{h}^2\mathcal{E}\mp 12\tilde{h}\mathcal{E}^2 +8\mathcal{E}^3=0
\end{equation}  
with solutions $\mathcal{E}_\pm=\tilde{h} [\pm \frac{1}{2};\mp \frac{3}{2};\pm \frac{5}{2}]$. They represent dressed states energies of an isotropic spin 5/2 with Rabi frequency:
\begin{equation} \label{eq:A7}
F_{5/2}^{iso}=F_{1/2}=\co h_{mw}.  
\end{equation}

\bibliographystyle{revtex4-1}

\end{document}